\documentclass[aps,pre,twocolumn,superscriptaddress,amsmath,amssymb,floatfix]{revtex4-1} 

\usepackage{graphicx}
\usepackage[usenames]{color} 
\usepackage{comment} 
\usepackage{bm}
\usepackage{amsmath}

\newcounter{abc}

\usepackage{graphicx}
\usepackage{amsmath}

\begin{document}

\title{Self-assembly and crystallisation of indented colloids at a planar wall}

\author{Douglas J. Ashton}
\affiliation{Department of Physics, University of Bath, Bath BA2 7AY}
\author{Samantha J. Ivell}
\affiliation{Department of Chemistry, Physical and Theoretical Chemistry Laboratory, University of Oxford, South Parks Road, Oxford, OX1 3QZ}
\author{Roel P. A. Dullens}
\affiliation{Department of Chemistry, Physical and Theoretical Chemistry Laboratory, University of Oxford, South Parks Road, Oxford, OX1 3QZ}
\author{Robert L. Jack}
\affiliation{Department of Physics, University of Bath, Bath BA2 7AY}
\author{Nigel B. Wilding}
\affiliation{Department of Physics, University of Bath, Bath BA2 7AY}
\author{Dirk G. A. L. Aarts}
\affiliation{Department of Chemistry, Physical and Theoretical Chemistry Laboratory, University of Oxford, South Parks Road, Oxford, OX1 3QZ}

\begin{abstract}
We report experimental and simulation studies of the structure of a
monolayer of indented (``lock and key'') colloids, on a planar
surface.  On adding a non-absorbing polymer with prescribed radius and
volume fraction, depletion interactions are induced between the
colloids, with controlled range and strength.  For spherical
particles, this leads to crystallisation, but the indented colloids
crystallise less easily than spheres, in both simulation and
experiment.  Nevertheless, simulations show that indented colloids do
form plastic (rotator) crystals.  We discuss the conditions under
which this occurs, and the possibilities of lower-symmetry crystal
states.  We also comment on the
kinetic accessibility of these states.
\end{abstract}

\maketitle

\section{Introduction}

Self-assembly of colloidal systems is a fast-moving area of current soft-matter reseach
-- the synthesis of novel micron-sized particles with controllable anisotropic interactions has allowed
the assembly of clusters, ``colloidal molecules'' and unusual crystals~\cite{Glotzer2007,Blaaderen2003,Sacanna2010,Rossi2011,Chen2011,Sacanna2012}.  
In some cases, anisotropic interactions can be realised by chemical patterning or ``patches'' on the surface of colloidal
particles~\cite{Kraft2009,Pawar2010,Chen2011,Groschel2013}.  Alternatively, the interplay between particle shape and a depletion interaction can drive self-assembly~\cite{Zhao2007,Odriozola2008,Sacanna2010,Rossi2011,Kraft2012,Ashton2013,Ortiz2014}.
One advantage of the depletion interaction is that the strength, range, and specificity of the interaction
can be tuned by the properties of the depletant molecules (usually a non-adsorbing polymer), 
so even a single synthesis of a colloidal system already allows access to a wide range of assembly conditions.

An extra degree of control over self-assembly can be achieved if it takes place under confinement,
or at a surface.  This effect is particularly strong for anisotropic particles in the presence of depletant,
because surfaces can affect the packing of the colloids~\cite{Kaplan1994,Riley2010,ivell2013}, and depletion forces also acquire an extra orientationally
dependent component, associated with binding of the colloids to the surface.  These surface effects lead to
new possibilities for controllable self-assembly.

Here, we consider ``indented'' or ``lock-and-key'' colloids~\cite{Sacanna2010,Sacanna2013}.  These particles
have spherical indentations in their surface, in which similar colloids can fit tightly.  This mode of binding is
favoured by the depletion effect~\cite{Odriozola2008,Ashton2013,Odriozola2013}.  
We have performed experiments on these particles, and simulated them by a Monte Carlo method.  
The particles are localised by gravity, against the (bottom) hard wall of their container.  
The experiments show clearly
that the indentations on the particles act to frustrate crystallisation,
and this fact is borne out by the simulations.  Similar frustration has been observed before in
systems of anisotropic particles~\cite{Dullens2006,Dullens2007,Rice2012} but in those cases there was significant polydispersity in the particle shapes:
here, the experimental
polydispersity is small, and the computational system is strictly monodisperse.

Moreover, our simulations also demonstrate
crystallisation of indented particles, in cases where depletion interactions are not too strong or short-ranged.
We propose a low-symmetry crystal state that we expect to be stable when depletion interactions are strong, 
but we find that its self-assembly is frustrated by kinetic factors.  Overall, our results illustrate the rich behaviour that can arise from a combination of particle shape, depletion interactions, and interfacial effects.

\section{Methods}

\subsection{Experiments}

The colloidal particles were synthesized following protocols reported in the literature \cite{Sacanna2010,Rossi2011}. Some modifications were made in order to incorporate a fluorescent dye, rhodamine B isothiocyanate (RITC), for confocal studies\cite{ivell2013}. The indented and the spherical particles were made following similar steps:  In short, both sets of particles were synthesized from 3-methacryloxypropyl trimethoxysilane (TPM; sourced from Polysciences and used as received). The TPM was hydrolysed under basic conditions, forming a cloudy dispersion of small droplets, which were subsequently grown to the desired size by adding more hydrolysed TPM. Modified RITC was added for fluorescence. At this stage the droplets were either cross-linked from the outside inwards, by adding potassium persulfate (Sigma Aldrich), leading to uniformly indented particles, or, from the inside outwards, by adding azobisisobutyronitrile (BDH Laboratories), which resulted in spherical particles. Both the indented and the spherical particles were made from the same batch of hydrolysed TPM droplets, allowing for a straightforward comparison in the behaviour of both systems. Both sets of particles had a diameter of $\sigma=2.56\mu$m and a polydispersity of $4\%$, as measured by static light scattering. The indented particles had a dimple of width $\sim 1.3\mu$m and depth $\sim 200$ nm, as determined by scanning electron microscopy and
atomic force microscopy, respectively.  Details of the above synthesis method can be found in reference \cite{ivell2013}.

The particles were suspended in various aqueous polymer solutions, where the polymers induce a depletion interaction. Specifically, we used solutions of xanthan (molecular weight $M_w = 3\times10^6$ g mol$^{-1}$ and calculated radius of gyration $R_G = 222$ nm \cite{koenderink03}) and of poly(ethylene oxide) (PEO; $M_w = 1\times10^6$ g mol$^{-1}$ and calculated $R_G = 57$ nm \cite{shang08}), with added salt (0.1M NaCl) to screen the double layer repulsion between the particles. We will focus here on results for PEO; as explained below, the smaller polymer leads to a more specific, directed depletion interaction. 

Multiple solutions of colloid-polymer mixtures were prepared in order to study the effect of the concentration of both colloidal particles and polymer in the system. 
The gravitational length of the colloidal particles was $0.2\mu {\rm m}$, so they form a colloidal monolayer at the base of the container, with negligible out-of-plane fluctuations. Any fluctuations away from the wall are also suppressed by the depletion interaction between particles and wall. Note that the polymers were not affected by gravity due to their negligible buoyant mass.
Due to the sedimentation of the particles, the variables of interest were the colloidal area fraction
and the polymer volume fraction. Here, we will focus on results for a system with a colloidal area fraction of $\phi_c \approx 0.5$, and with polymer volume fractions of $\phi_p \approx 0.5$.

Samples were left to sediment completely, which took approximately two hours.  
The resulting monolayer
 was then observed using a Zeiss LSM 5 Exciter confocal microscope fitted with a 63$\times$— oil immersion
objective. For each sample, movies of 150 frames were recorded, where
each frame was 51.2 $\mu$m $\times$ 51.2 $\mu$m in size. The 2D coordinates of each
particle were found using a tracking routine.

\subsection{Simulations}

\begin{figure}
\begin{center}
\includegraphics[width=8.5cm]{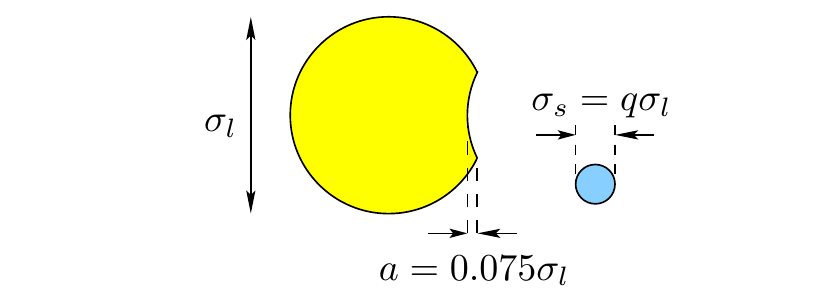}
\caption{Simulation model for the indented colloids.  The size ratio 
between colloidal particles (yellow) and ideal depletant particles (blue) is $q$.  The indentation
depth is fixed at $a=0.075\sigma_l$ throughout this work, for consistency with the experimental system.}
\label{fig:model}
\end{center}
\end{figure}

We use Monte Carlo (MC) simulations to calculate the equilibrium properties of indented colloids, in the presence
of depletion interactions, confined close to a hard wall.
%
We model these particles by taking spheres of diameter $\sigma_l$, 
and cutting away a volume that corresponds to its intersection with a second sphere of the same diameter.  
The resulting situation is shown in Fig.~\ref{fig:model}: the
distance between the centres of the spheres is $d_c=0.85\sigma_l$, so the depth of the indentation is  $0.075 \sigma_l$,
comparable with the experimental case (approximately $0.078\sigma$).  
The depletant is modelled as a fluid of small spheres of diameter $\sigma_s$,
which we parameterise by the size ratio $q=\sigma_s/\sigma_l$.  These spheres are an `ideal' depletant~\cite{Asakura1954}: they interact
with the colloids as if they were hard particles, but they do not interact with each other.  The chemical potential of the depletant
is adjusted so that their volume fraction in a system without colloids would be $\eta_s$.

We place $N=100$ colloidal particles in a cuboidal box of dimensions
$L_x=L_y=12.5 \sigma_l$, $L_z=2.5 \sigma_l$. Structureless hard walls
are placed at $z=0$ and $z=L_z$ and gravity acts in the
negative $z$ direction. 
These parameters give an area fraction for the
colloids of $\phi_c \approx 0.5$. 
The gravitational length associated with the colloidal particles is $0.1\sigma_l$,
comparable with experiments.
Given this gravitational length, the height of the box is sufficient
to prevent any effects from the upper wall.

In order to obtain good sampling in this system, we use grand-canonical insertion/deletion moves for the depletant,
combined with the geometric cluster algorithm (GCA)~\cite{Liu2004} as described in \cite{Ashton2013}. The
complete move set includes 
(i) grand canonical moves for the small particles,
(ii) standard displacement and rotation moves of the colloids,
(iii) GCA ``biased pivot'' moves, where the pivot is placed close to the centre of the
particle to move it a small distance, 
(iv) GCA plane moves where a particle can rotate by an arbitrary amount,
(v) GCA ``biased plane'' moves where the particle's director nearly lies in the reflection plane resulting in a small
rotation, and
(vi) combinations of pivot and plane GCA moves.
At the beginning of the simulations, the colloids are first equilibrated without depletant to ensure they drop to
the bottom of the box, after which grand canonical insertion/deletion MC moves for the depletant are turned on

The aim of the GCA is to move clusters of particles together, since otherwise the large number of depletant
particles in the system tend to obstruct the movement of the colloids.  However, for computational efficiency,
it is sometimes convenient to restrict the size of the cluster being moved, since moves of larger clusters require
greater computational effort.  Therefore, in
each Monte Carlo sweep we perform $N$ moves where the cluster can contain only one
indented colloid (but with unlimited depletant particles), as well as one move where the cluster
size is unlimited. We perform these moves in
3-dimensions to allow vertical movement of the colloids, but we also perform updaes where colloids
move only in the $xy$-plane, which allows efficient relaxation in  two dimensions.  Our
implementation of the GCA with gravity includes an extra step. As each
particle move is proposed, we test for acceptance using a Metropolis criterion,
based on the change in gravitational potential energy. 
If any colloidal particle fails this test the whole move is rejected.

\subsection{Depletion interactions}

\begin{figure}
\begin{center}
\includegraphics[width=7.5cm]{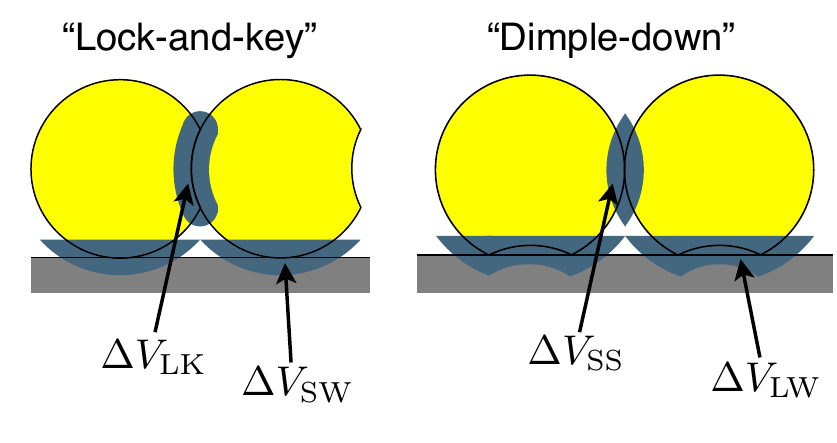}
\caption{Two binding modes for the depletion interaction next to a hard wall (grey).  The blue regions indicate the excess free volume $\Delta V$
that becomes accessible
to depletant particles when the colloids bind as shown.   The strength of the depletion interaction
is proportional to these volumes.  In ``lock-and-key"
binding, one lock sits within the indentation of another, while in ``dimple-down" binding, the indentation points towards the wall.  
Numerical and exact geometrical calculations indicate that $\Delta V_{\rm LK}+\Delta V_{\rm SW} > \Delta V_{\rm SS}+\Delta V_{\rm LW}$,
which means that lock-and-key binding is typically the dominant binding mechanism.}
\label{fig:depletion}
\end{center}
\end{figure}

The structures that are formed by these indented colloids are controlled by depletion interactions.
These interactions arise because if two 
colloidal particles come close to each other, the volume accessible to the depletant particles is increased.  This effect increases the
total entropy of the system, and the result is an attractive  force between the colloids. There are also
attractive depletion forces between the colloids and the hard walls of the system.

The decrease in free energy (gain in entropy) associated with various configurations of the colloids can
be estimated by a geometrical argument.  The most relevant cases are illustrated in Fig.~\ref{fig:depletion}: they are (a)
``lock-and-key'' binding, where one colloidal particle sits
within the indentation of another colloid; (b) ``dimple-down'' binding, where the indentation on the colloidal particle 
points towards a nearby hard wall. The sizes of the shaded
areas in Fig.~\ref{fig:depletion} indicate the volume released to the depletant when the colloids bind -- the larger the relevant volume, the stronger
is the attractive depletion force.  These volumes can be calculated geometrically.  

For the purposes
of this work, we emphasize two key points.  First, the ``lock-and-key'' binding mechanism is associated with the strongest
depletion force, while ``dimple-down'' binding is rather weaker.  The relevant volumes that are released to the depletant are
illustrated in Fig.~\ref{fig:depletion}: we label these as ``lock-and-key'' ($\Delta V_{\rm LK}$); ``lock-and-wall'' ($\Delta V_{\rm LW}$); 
``sphere-to-sphere'' ($\Delta V_{\rm SS})$ and ``sphere-to-wall'' ($\Delta V_{\rm SW}$).  When adding a particle to an existing
cluster, ``lock-and-key'' binding is favoured over ``dimple-down'' binding if $\Delta V_{\rm LK}+\Delta V_{\rm SW} > \Delta V_{\rm SS}+\Delta V_{\rm LW}$.
This inequality is satisfied for the particles considered here.
Second, the differences in bond strength for the different binding mechanisms are larger when the
depletant particles are smaller (small-$q$).  It may be useful to think of smaller-$q$ corresponding to ``more specific'' depletion
interactions, both in terms of the relative strength of the different bonds, and in terms of the range of the depletion interaction.

\section{Results}
\subsection{Structures of spherical and indented colloids}

\begin{figure}
\begin{center}
\includegraphics[width=8.5cm]{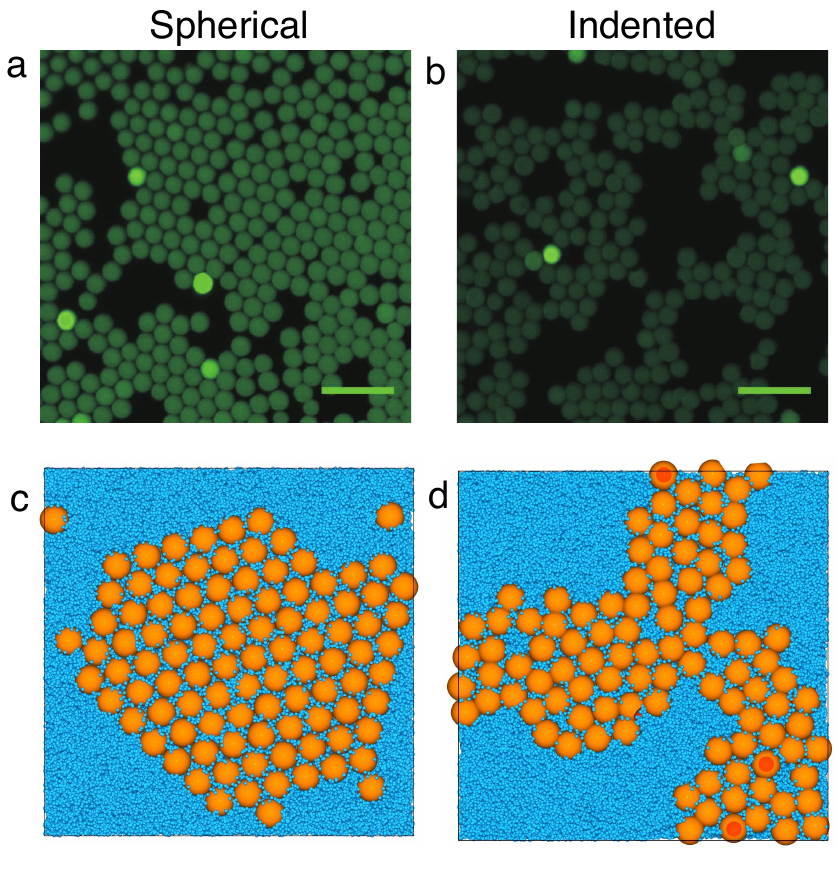}
\caption{Typical configurations of spherical and indented colloids in 
experiment (a,b) and simulation (c,d).  The colloid area fraction
is $\phi_{\rm c}\approx 0.5$ in all cases.  In the experiments, the
the depletant (not visible) is PEO at volume fraction $\phi_p=0.5$, in a 0.1M salt solution.  Scale bars are $10\mu{\rm m}$.
The simulation results are at size ratio $q=0.125$, which gives qualitative
agreement with these experiments.  The depletant volume fraction is $\eta_{\rm s}=0.36$,
which was chosen to maximise the crystallinity of the indented sphere system (see Fig.~\ref{fig:chi6} below).
The simulation snapshots are visualised from below: in (d), there are several particles whose
indentations (coloured red) are oriented towards the wall (``dimple-down'' configuration). 
}
\label{fig:snapspherelock}
\end{center}
\end{figure}

\begin{figure}
\begin{center}
\includegraphics[width=8.5cm]{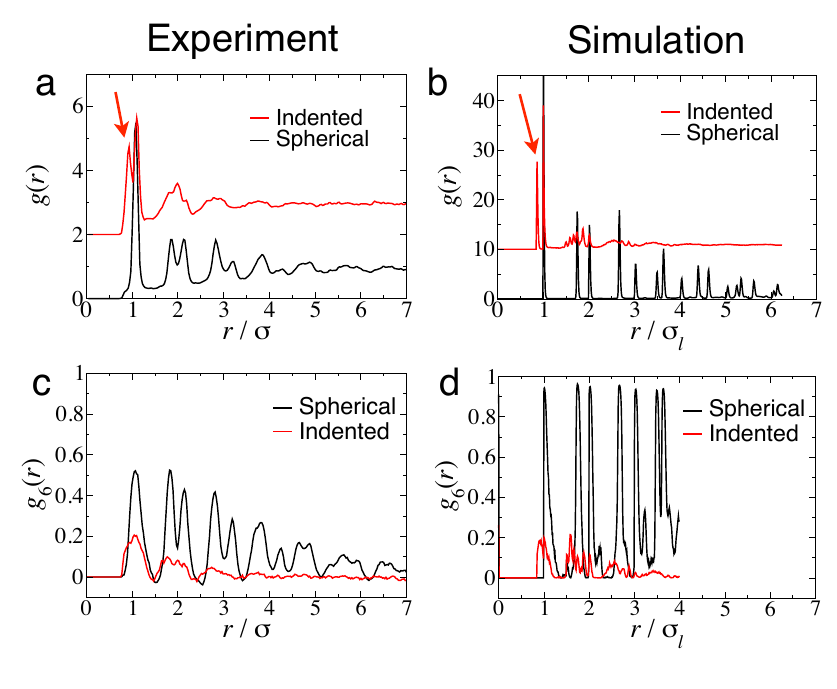}
\caption{%
(a,b) Radial distribution function $g(r)$, showing the extent of positional order in 
experiments and simulations, under the same conditions as Fig.~\ref{fig:snapspherelock}.  Arrows indicate the peaks
associated with``lock-and-key'' binding, at $r=0.85\sigma_l$.
(c,d) Correlation function $g_6(r)$, showing the extent of bond-orientational order.
Compared with the spheres, the indented colloids show less crystalline and orientational order, in all cases.  Also,
the simulations assemble into more ordered structures, compared with the experiments.
}
\label{fig:grspheredimp}
\end{center}
\end{figure}

In Fig.~\ref{fig:snapspherelock}, we show self-assembled structures in the experimental
and computational systems.  The colloid area fraction is $\phi_c\approx 0.5$ in all cases,
and there is qualitative agreement between simulation and experiment -- while the spherical particles 
readily crystallise, the indented spheres form large disordered clusters, reminiscent of vapour-liquid coexistence.  In the simulations, the size ratio between colloid and depletant is $q=0.125$, larger than
the experimental case; the volume fractions of depletant used in simulation are also lower.
This indicates that the depletion interaction in the experiments is weaker than that predicted by the idealised
model used in the simulations, which we attribute primarily to
non-ideal colloid-depletant interactions \cite{bookremcohenk,Germain2004}
and (for the indented colloids) an indentation that does not match the perfect spherical shape used 
in the simulations.  However, the qualitative features of the experiments are well captured by the simulations. 
The dependence of the simulation results on depletant volume fraction is discussed in Section~\ref{subsec:dependence}
below.  We also performed experiments using a larger polymer (xanthan).  The results are qualitatively similar,
in that spheres crystallise more readily than indented colloids.

To quantify the extent of crystallinity and bond-orientational order, 
we measured the two-dimensional radial distribution function
of the colloids $g(r)$, as well as a measure of orientational order $g_6(r)$.  To
define $g_6(r)$, we first introduce a local bond-order parameter for particle $i$:
\begin{equation} \label{eq:phi6}
\phi_6(i) = \frac{1}{n} \sum_j e^{{\rm i} 6 \theta_{ij}}
\end{equation}
where the sum runs over neighbours $j$ of particle $i$, the number of these neighbours is $n$ and 
$\theta_{ij}$ is the angle between $\vec{r}_{ij}=\vec{r}_j-\vec{r}_i$ and an arbitrary axis.
In systems with bond orientational order, the complex numbers $\phi_6(i)$ have significant interparticle correlations.
In particular, 
\begin{equation}
g_6(r) = \frac{ \langle  \phi_6(i)  \phi_6^*(j) \delta(r - |\vec{r}_{ij}|) \rangle}{ \langle \delta(r - |\vec{r}_{ij}|) \rangle}
\end{equation}
measures order at distance $r$, with $g_6(r)=0$ if the system has no orientational order at length scale $r$, while
$g_6(r)=1$ if the
system has perfect bond orientational order.
It is also useful to define a ``susceptibility'', which is related to a spatial integral of $g_6$, and gives the approximate
size of orientationally-ordered domains in the system:
\begin{equation} \label{eq:chi6}
\chi_6 = \frac{1}{N}  \left\langle \Big| \sum_{i}  \phi_6(i) \Big|^2 \right\rangle .
\end{equation}

Results for positional order [$g(r)$] and orientational order [$g_6(r)$] are shown in Fig.~\ref{fig:grspheredimp}.  Several
features are notable.  For $g(r)$, the simulation results for spheres show several sharp peaks, consistent with the almost-perfect
crystal state shown in Fig.~\ref{fig:snapspherelock}.  The corresponding data for the indented particles lacks the sharp peaks
at $r>\sigma_l$, but there is a new peak that appears at $r\approx 0.85\sigma_l$.  This peak corresponds to the onset of
lock-and-key binding for the indented particles: there is a strong depletion force that favours the state where one colloid sits
snugly in the indentation of another.  The experimental data show the same qualitative behaviour: the spheres show a splitting in the second peak of $g(r)$, as expected for crystalline states, and there are other peaks in $g(r)$ for $r>2\sigma$, consistent
with crystalline order.  It is notable that the peaks in $g(r)$ are much less sharp for the experiment as compared with the simulation: we attribute this to polydispersity among the colloids, and the inherent uncertainty in capturing particle positions from microscope images.  Turning to the experimental data for the indented colloids, one observes a suppression of long-ranged positional order,
and the appearance of a new lock-and-key binding peak at $r\approx 0.85\sigma$.

The orientational correlation function $g_6(r)$ reinforces this overall picture:  the spheres show clear evidence for long-ranged
bond-orientational order, as expected for crystalline states, while the indented particles have shorter-ranged correlations, consistent with the suppression of crystalline order by the lock-and-key bonds.  As was clear from $g(r)$, the lock-and-key bonds are shorter than the usual sphere-sphere bonds: the presence of two competing length scales for binding acts to suppress the crystalline state.

Given that both simulation and experimental results indicate that crystallisation is suppressed by indentations in the colloidal particles, we now use simulation studies to investigate this effect in more detail.  In particular, we concentrate on the effects of varying depletant size and volume fraction on the indented colloids: by changing the range of the depletion interaction, we are able to tune the system from a state where crystallisation is rapid into a state where crystallisation is frustrated by lock-key binding.

\subsection{Dependence on depletant size and volume fraction} 
\label{subsec:dependence}

\begin{figure}
\begin{center}
\includegraphics[width=8.5cm]{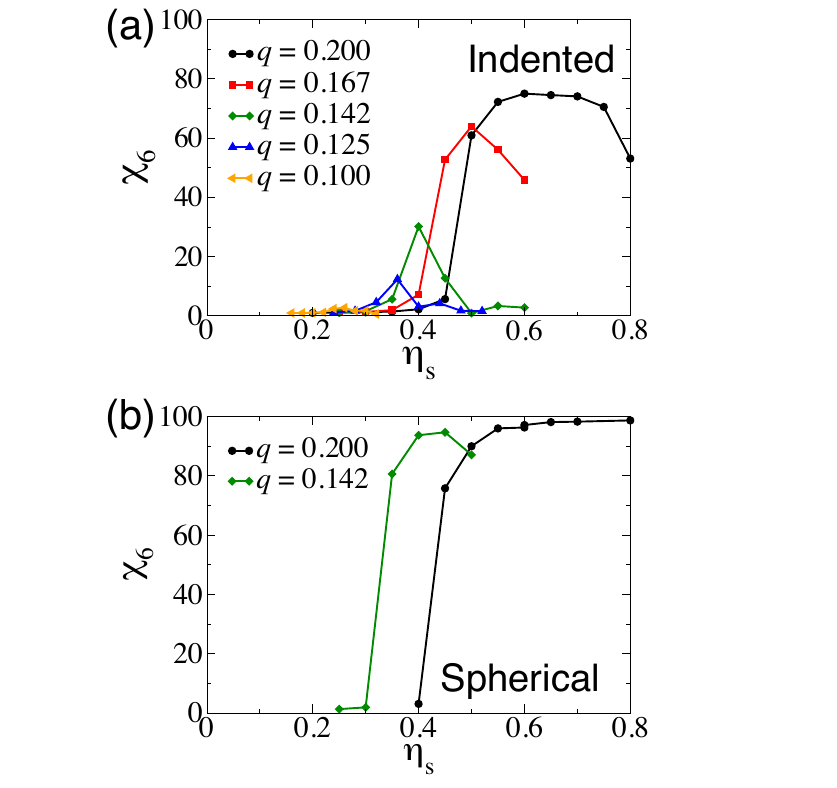}

\caption{Measurements of global bond-orientational order ($\chi_6$) obtained from
  simulations, as a function of depletant volume fraction $\eta_{\rm
    s}$, for various sizes ratios $q$.  Comparing spheres and indented
  colloids, the spheres crystallise more readily and over a larger
  range of $\eta_{\rm s}$, although there is evidence for kinetic
  trapping effects at large $\eta_{\rm s}$ and small $q$.  For the indented
  colloids, the crystallinity is lower and the range of $\eta_{\rm s}$ in
  which significant crystallisation is observed is much narrower.%
}
\label{fig:chi6}
\end{center}
\end{figure}

\begin{figure}
\begin{center}
\includegraphics[width=8cm]{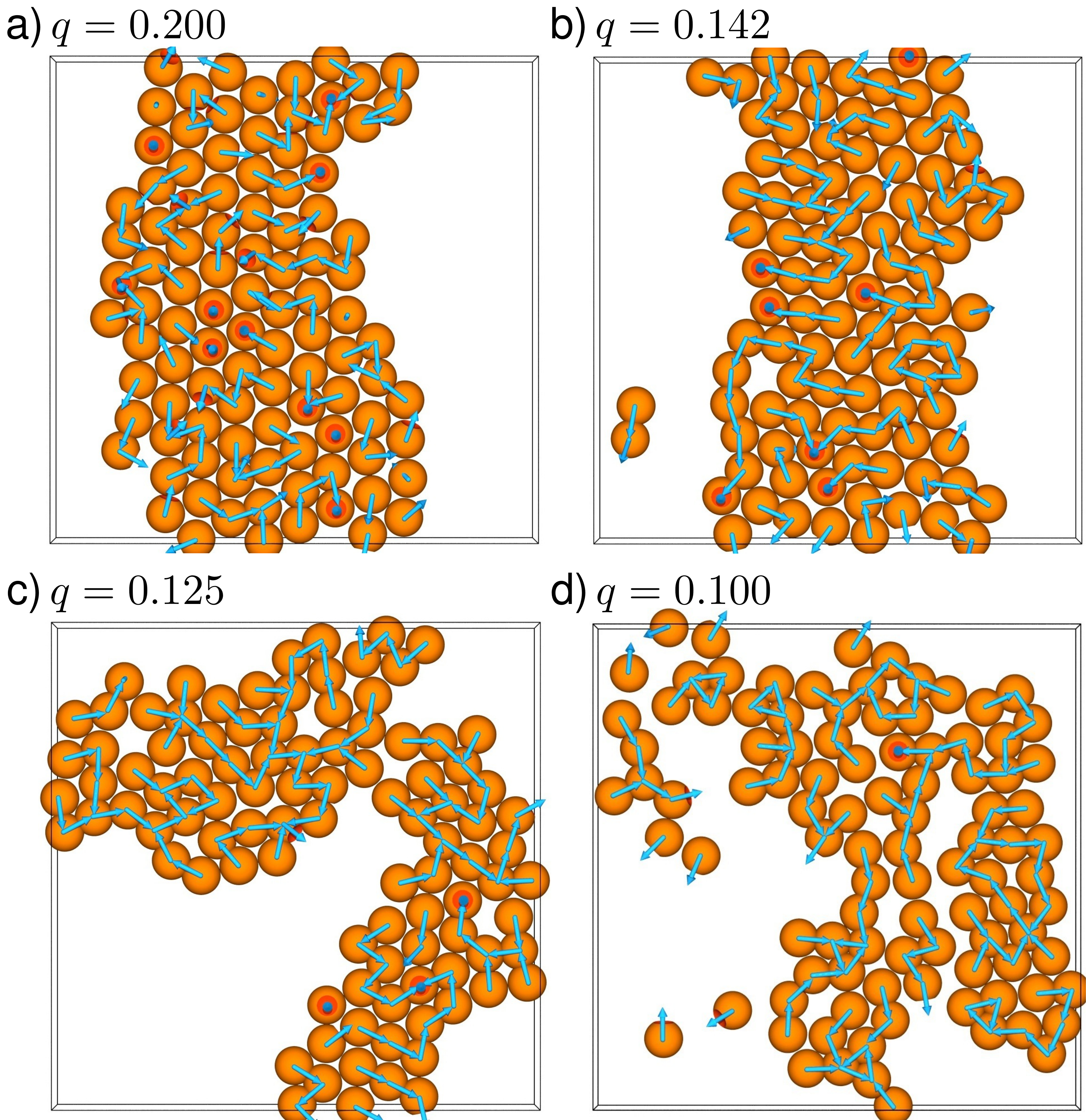}
\caption{Representative configurations from simulations of indented colloids, taken at the values of $\eta_{\rm s}$ at which $\chi_6$ is maximal in Fig.~\ref{fig:chi6}.   Colloidal orientations are illustrated using arrows pointing outward through the indentations.  For $q=0.2,0.142$, the colloidal orientations relax quickly on the simulation time scale, indicating that this is a plastic (rotator) crystal.  For smaller $q$, the orientational relaxation is much slower, consistent with the strong lock-and-key binding.}
\label{fig:snaps8}
\end{center}
\end{figure}

In Fig.~\ref{fig:chi6}, we collate results that show the total amount
of orientational order that forms in systems of indented colloids, on
varying the size and volume fraction of the depletant.  These results
are taken from long MC simulations which were initialised in a
disordered state and have ``equilibrated'' into a steady state -- this
state is not guaranteed to be the global free energy minimum of the
system but it is at least strongly metastable.  We discuss this
further in Section~\ref{subsec:kinetic} below: the essential point is
that if the simulations suffer from strong kinetic effects, it is
likely that similar effects will be observed in experiments.  So even
for systems that are not not fully equilibrated, we can expect the
metastable states found by simulation to be similar to those found in
experiment.

For a size ratio $q=0.2$ and $\eta\gtrsim0.45$, the system readily
crystallises, as is clear from the large values of $\chi_6$.  The maximum possible value of the
susceptibility $\chi_6$ in a finite system is equal to the 
total number of particles, $N=100$.  One observes $\chi_6<N$ either due to 
local distortions of the crystal lattice, or due to defects and
domain boundaries that disrupt the crystalline packing on relatively large length scales.
The results for $q=0.2$
indicate that the crystal domain size in the system is comparable with
the system size. Comparing with the spherical particles, the value of $\chi_6$ is suppressed: we
attribute this primarily to local structural disortion.  In this case, we can be confident that the system
has equilibrated in its thermodynamically stable state.  To reinforce
this message, Fig.~\ref{fig:snaps8} shows representative
configurations: for each size ratio $q$, we show a configuration at
the point that maximises $\chi_6$.  The particles are decorated by
arrows, which indicate their orientation: the arrow points out through
the centre of the indentation in the colloid.  At $q=0.2$, the system
is indeed crystalline: particles tend to be oriented with their
indentations pointing towards adjacent particles, but the range of the
interaction is long enough that the crystal is not frustrated.  There
are also a significant number of dimple-down particles, consistent
with this configuration being stabilised by the depletion interaction
(but not favoured as strongly as lock-and-key binding).  Indeed,
comparing with the data for spherical colloids shown in
Fig.~\ref{fig:chi6}(b) the indentations have rather little effect on
the crystallinity.  For the largest $\eta_s$, the crystallinity starts
to fall -- we attribute this primarily to kinetic effects: see
Sec.~\ref{subsec:kinetic} below.

As $q$ is decreased, the lock-and-key binding between colloids becomes stronger and shorter ranged, and this starts to disrupt the crystal formation.  The onset of crystalline order happens at a lower depletant volume fraction $\eta_s$, but the extent of crystalline order is strongly suppressed for $q<0.14$.  Fig.~\ref{fig:snaps8}(b) indicates the origin of this effect: the topology of the particle packing is close to a hexagonal crystal, but the shorter lock-and-key bonds leads to local distortions away from the perfect lattice.  (The ``crystal planes'' deviate from parallel straight lines, reducing orientational order.)  For larger $\eta_s$, the crystallinity decreases, which we again attribute primarily to kinetic effects.

\begin{figure}
\begin{center}
\includegraphics[width=7.5cm]{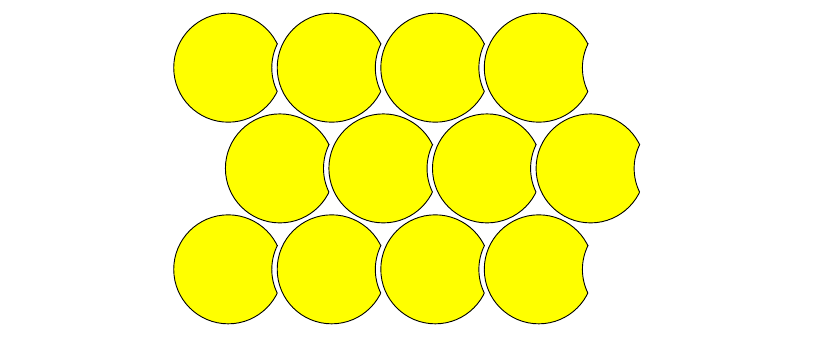}
\caption{Illustration (view from above) of the low-symmetry crystal that we expect to be stable when depletion
interactions are very strong.  The orientations of the colloids acquire long-ranged order, which breaks
the six-fold rotational symmetry of the original crystal -- the ``lock-and-key'' bonds parallel to the orientational order are shorter 
than the other interparticle bonds.}
\label{fig:column}
\end{center}
\end{figure}

For the smallest $q$, we see almost no evidence of crystallinity.  For these small depletant particles, the lock-and-key binding of the colloids is very strong, which leads to strong kinetic trapping~\cite{whitelam2009,Grant2011}.  Hence, self-assembly of ordered structures tends to be suppressed: this effect is apparent in both simulations and experiments.  Nevertheless, we can use theoretical arguments to obtain the expected fate of the system for small $q$.  We expect the strong lock-and-key binding in this regime to cause the formation of chains of colloids (``colloidal polymers''), which can also branch, leading (when bonds are strong) to large percolating clusters.  For small-$q$, the depletion interaction between the convex surfaces of the colloids is much weaker than the lock-and-key binding, so we expect a range of $\eta_s$ over which these branching chains dominate the system.  However, for larger $\eta_s$, the chains can reduce their free energy by clustering (or collapsing in on themselves) -- for the largest $\eta_s$ we expect the stable state to be a low-symmetry crystal such as that shown in Fig.~\ref{fig:column}.  It is interesting to note that neither the simulations nor the experiments solved the packing problem by turning the indentations towards the wall; although this would allow a perfect hexagonal packing, the loss in the depletion interaction compared to the lock-and-key binding is apparently too large for the systems studied here.

\subsection{Kinetic effects, and stability of low-symmetry crystals}
\label{subsec:kinetic}

\begin{figure}
\begin{center}
\includegraphics[width=8.4cm]{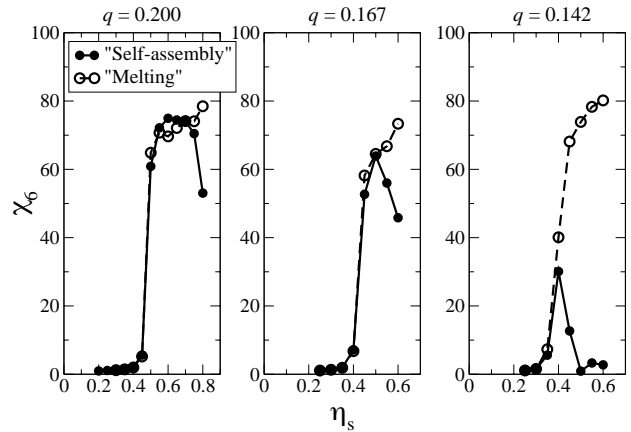}
\caption{Illustration of kinetic effects in MC simulations.  We show the extent of bond-orientational order, measured via $\chi_6$, obtained from the final states of
``self-assembly'' simulations (initialised in disordered states, closed symbols)
and ``melting'' simulations (initialised in the low-symmetry crystal shown in Fig.~\ref{fig:column}, open symbols). For $q=0.2$, the two initial conditions typically lead to 
similar results, showing that kinetic effects are small and the system readily equilibrates in a plastic (rotator) crystal.  For smaller $q$ and large $\eta_s$, the results
of the simulation depend strongly on the initial condition, indicating that kinetic effects are strong (see the main text for a full discussion).}
\label{fig:kinetic}
\end{center}
\end{figure}

As noted in the previous section, there are significant kinetic
trapping effects in these systems, due to the presence of strong
directional bonds.  For example, in simulations at increasing
depletant volume fraction $\eta_s$, one observes from
Fig.~\ref{fig:chi6} that the crystallinity of the assembled state
increases for small $\eta_s$, before decreasing again.  Such
non-monotonicity is familiar from other systems with strong kinetic
trapping~\cite{Hagan2006,whitelam2009,Grant2011}.  However, in this
system, such a non-monotonicity might also arise from purely
thermodynamic considerations -- the hexagonal crystal may be unstable
for very large $\eta_s$, due to the presence of two different length scales (shorter 
lock-and-key bonds and longer bonds between the convex surfaces of colloidal particles).

To distinguish between these kinetic and thermodynamic effects, we
performed simulations where the system was initialised in the
low-symmetry crystal state shown in Fig.~\ref{fig:column}, and then
relaxed using the GCA as before. This highly-ordered initial state was
chosen to contrast with the original ``self-assembly'' simulations
where the system was initialised in a disordered state.  For
relatively large $q$ and small $\eta_s$, the simulations gave the same
results, independent of the initial condition, indicating that the
system is fully equilibrated (see Fig.~\ref{fig:kinetic}).  We also
note that for $q=0.2$, the orientationally-ordered crystalline initial
state does relax to a plastic (``rotator'') crystal, providing further
evidence that this is the thermodynamically stable state under these
conditions. However, in the regime of large $\eta_s$, simulations
initialised as a crystal were observed to evolve to a state
whose degree of crystallinity increased with $\eta_s$. This contrasts
with the non-monotonic behaviour observed in the ``self assembly''
simulations and indicates that the crystal state is at least
metastable under these high-$\eta_s$ conditions.

In Fig.~\ref{fig:kinetic}, we expect the ``self-assembly'' and
``melting'' simulations to give approximate upper and lower bounds on
the true equilibrium value of $\chi_6$.  Since the extent of orientational
order is already significant at the points where kinetic trapping sets
in, we believe that states with bond-orientational order are likely to become
increasingly stable as $\eta_s$ increases, and therefore that the
equilibrium value of $\chi_6$ is likely to be monotonic in $\eta_s$.
In this case, the non-monotonicity of the crystallinity as a function
of $\eta_s$ in Fig.~\ref{fig:chi6}(b) would be primarily a kinetic
effect.  However, we cannot rule out a thermodynamic component of this
non-monotonicity -- this would require a detailed numerical study of
the phase behaviour of this system, which is beyond the scope of this
work.

\section{Conclusions}

We have shown how adding indentations to colloidal spheres acts to
frustrate the self-assembly of two-dimensional crystals.  However, by
varying properties of the depletant, we find that crystallisation can
still be observed (in simulations) when the depletant particles are
not too small, and the depletion forces not too strong.  We emphasise
that it requires the depletant properties to be tuned quite
accurately, a situation familiar from other self-assembly processes;
indeed, crystallisation of indented colloids was not observed in the
experiments.  Possibilities for `tuning' systems towards a narrow
regime of effective assembly has been discussed recently, but this
remains a challenging problem~\cite{Jack2007,Jankowski2012,Klotsa2013}.

We have proposed that the low-symmetry crystal shown in
Fig.~\ref{fig:column} should be thermodynamically stable in the limit
where depletion forces are strong and short-ranged.  However, while
this phase was stable within our simulations, its self-assembly was
not observed, presumably due to kinetic effects.  The behaviour of the
system in this regime remains an interesting area for study -- this
might be achieved by using an effective potential to describe the
colloid-colloid interactions, instead of treating the depletant
explicitly, as in this study.

Overall, we have emphasised the range of behaviour that can be observed through a combination
of anisotropic colloid shape, depletion interactions, and the effects of confinement close to a wall.  As it becomes
possible to make colloidal particles with increasingly complex and well-controlled shapes, we argue that self-assembly
in these kinds of setting represents a promising avenue for future progress.  However, the frustration of self-assembly by kinetic factors that are observed even in this relatively simple system emphasizes that the complex dynamics of self-assembly
must be borne in mind when designing these processes.  Alternatively, one might turn the kinetic frustration of
crystallisation into an advantage: systems where crystallisation is suppressed due to a non-spherical colloidal shape might
have potential application as colloidal glass-formers~\cite{Rice2012}.

\begin{acknowledgments}
We are grateful to the EPRSC for  support for DJA and NBW through grant EP/I036192/1; a PhD studentship (for SI); and support for RLJ through grant 
EP/I003797/1.
\end{acknowledgments}

\bibliography{lock-paper}

\end{document}